\documentclass[aps,prl,twocolumn,groupedaddress]{revtex4-1}
\usepackage{accents}
\usepackage{graphicx}
\usepackage{amssymb}

\begin{document}

\title{Inflation with very small tensor-to-scalar ratio}


\author{Koichi Hirano}
\email[]{k\_hirano@tsuru.ac.jp}
\affiliation{Department of Teacher Education, Tsuru University, 3-8-1, Tahara, Tsuru, Yamanashi, 402-8555, Japan}


\date{\today}

\begin{abstract}
We have investigated inflation models that predict a very small value of the tensor-to-scalar ratio, $r$. The spectral index $n_s$, and the tensor-to-scalar ratio $r$, are strictly constrained by the Planck data. $n_s$ and $r$ are sensitive to the shape and magnitude of the inflaton potential, respectively.The constraints by the Planck 2018 data combined with other cosmological observations are compared with the predictions from the inflation models regarding $n_s$ and $r$. Furthermore, we discuss the comparison of future tensor-to-scalar ratio data with predictions from the inflation models with a focus on part of the quantum fluctuation origin.\end{abstract}


\maketitle

\section{Introduction}
Big-bang cosmology can account well for the observational results in cosmology, such as Hubble expansion, the intensity of the cosmic microwave background (CMB) radiation, and the primordial abundance of light elements. However, it faces a horizon problem, the flatness problem and monopole problem, and it does not explain the origin of primordial CMB fluctuations \cite{Smooth1992}.

Therefore, an epoch of accelerated expansion in the early universe, i.e., inflation \cite{Guth1981,Sato1981a,Sato1981b,Starobinsky1980}, must be considered to solve these fundamental problems. Quantum fluctuations generated during inflation become the seeds of density perturbations observed in the CMB anisotropies, and the origin of the large-scale structure. Understanding of the inflation mechanism is thus very important to advance physics beyond the standard model.

Many inflation models have been suggested. These different models can be broadly categorized in the following way. The first class is the large field model (type I), in which the inflaton field is initially large and rolls down toward the potential minimum at smaller $\phi$. Chaotic inflation is one of the representative models of this type \cite{Linde1983}. The second class is the small field model (type II), in which the initial value of the inflaton is small and it slowly evolves toward the potential minimum at larger $\phi$. New inflation and natural inflation are examples of this class \cite{Linde1982,Albrecht1982,Freese1990,Adams1993}.

The third class is the double inflation model (type III) \cite{Linde1994,Copeland1994}, in which inflation typically ends by the phase transition caused by the existence of a second scalar field (or by the second phase of inflation subsequent to the phase transition).
There are models of inflation that cannot be classified into the above three types.

The primordial density perturbations are characterized by the spectral index $n_s$, and the tensor-to-scalar ratio, $r$. $n_s$ and $r$ are generally sensitive to the shape and magnitude of the inflaton potential, respectively, and are strictly constrained by the Planck data combined with other cosmological data \cite{Akrami2018}.

In this paper, we investigate those inflation models that predict the very small value of $r$. The constraints by the Planck 2018 data combined with other cosmological data are compared with the predictions from the inflation models with respect to $n_s$ and $r$. Furthermore, we discuss the comparison of future tensor-to-scalar ratio data with the predictions from the inflation models with a focus on part of the quantum fluctuation origin.

This paper is organized as follows. In the next section, we present inflation models that predict the very small value of $r$. In Section III, we compare the constraints by the Planck 2018 data combined with other cosmological data with the predictions from the inflation models. In Section IV, we discuss the comparison of tensor-to-scalar ratio data by future observations with the predictions from the inflation models. Finally, a summary is given in Section V.

\section{Models}

We begin with the most general scalar-tensor theories with second-order equations of motion
\begin{equation}
\mathcal{L} = \sum^5_{i=2}\mathcal{L}_i, \label{eq:lag_horn0}
\end{equation}
where
\begin{eqnarray}
\mathcal{L}_2 & = & K(\phi,X), \\
\mathcal{L}_3 & = & -G_3(\phi,X)\Box\phi,  \\
\mathcal{L}_4 & = & G_4(\phi,X)R+G_{4,X}[(\Box\phi)^2-(\nabla_\mu\nabla_\nu\phi)(\nabla^\mu\nabla^\nu\phi)], \\
\mathcal{L}_5 & = & G_5(\phi,X)G_{\mu\nu}(\nabla^\mu\nabla^\nu\phi)-\frac{1}{6}G_{5,X}[(\Box\phi)^3 \nonumber \\
& & -3(\Box\phi)(\nabla_\mu\nabla_\nu\phi)(\nabla^\mu\nabla^\nu\phi)+2(\nabla^\mu\nabla_\alpha\phi)(\nabla^\alpha\nabla_\beta\phi)(\nabla^\beta\nabla_\mu\phi)]. \nonumber \\ \label{eq:lag_horn5}
\end{eqnarray}
Here, $K$ and $G_i$ ($i = 3, 4, 5$) are functions of an inflaton field $\phi$, and its kinetic energy $X = -\partial^\mu\phi\partial_\mu\phi/2$ with the partial derivatives $G_{i,X} \equiv \partial G_i/\partial X$. $R$ is the Ricci scalar, and $G_{\mu\nu}$ is the Einstein tensor. This Lagrangian function was first proposed by Horndeski in a different notation \cite{Horndeski1974}. The Lagrangian (Eqs. (\ref{eq:lag_horn0})-(\ref{eq:lag_horn5})) is equal in content to that derived by Horndeski \cite{Deffayet2011,Kobayashi2011}. The total action we are going to study is given by:
\begin{equation}
S = \int d^4x\sqrt{-g}\mathcal{L},
\end{equation}
where $g$ is a determinant of the metric $g_{\mu\nu}$.

In this work, we consider potential-driven slow-roll inflation. This corresponds to a case where the functions in the Horndeski theory of Eqs. (\ref{eq:lag_horn0})-(\ref{eq:lag_horn5}) are expressed as:
\begin{eqnarray}
K(\phi,X) & = & X-V(\phi), \\
G_3(\phi,X) & = & 0, \\
G_4(\phi,X) & = & \frac{M_{\rm pl}^2}{2}, \\
G_5(\phi,X) & = & 0,
\end{eqnarray}
where $V(\phi)$ is the inflaton potential, and $M_{\rm pl}$ is the reduced Planck mass associated with Newton's gravitational constant by $M_{\rm pl} = 1/\sqrt{8\pi G}$.

For the flat Friedmann-Lema$\hat{\i}$tre-Robertson-Walker (FLRW) metric described by the line element $ds^2 = -dt^2+a^2(t)\delta_{ij}dx^idx^j$, the Friedmann equation and the inflaton field equation of motion are given, respectively, as:
\begin{equation}
3M^2_{\rm pl}H^2=\frac{1}{2}\dot{\phi}^2+V(\phi) , \label{eq:Friedmann}
\end{equation}
\begin{equation}
\ddot{\phi}+3H\dot{\phi}=-V,_\phi , \label{eq:FieldMotion}
\end{equation}
where $H=\dot{a}/a$ is the Hubble parameter (the dot represents a derivative with respect to $t$). The notation $V,_\phi\equiv\partial V(\phi)/\partial\phi$ is adopted here.

Inflation can be induced in a regime where the slow-roll parameter $\epsilon\equiv-\dot{H}/H^2$ is much smaller than 1. Using the slow-roll parameter $\epsilon$, parameter $\eta$ is defined as:
\begin{equation}
\eta\equiv\frac{\dot{\epsilon}}{H\epsilon}.
\end{equation}

The number of e-foldings is set as $N(t) = \ln{a(t_f)}/a(t)$, where $a(t)$ and $a(t_f)$ are the scale factors at time $t$ during inflation, and at the end of inflation, respectively. From the relation $dN/dt = -H(t)$, N(t) can also be described as:
\begin{equation}
N(t) = -\int_{t_f}^t H(\tilde{t})d\tilde{t}, \label{eq:efoldings}
\end{equation}
and there is the condition $\epsilon(t_f)=1$ for $t_f$. The number of e-foldings when the perturbations related to the CMB temperature anisotropies get across the Hubble radius is typically $50 < N < 60$ \cite{Ade2014}.

Using the slow-roll approximations ${\dot{\phi}}^2/2\ll V$ and $|\ddot{\phi}|\ll |3H\dot{\phi}|$, Eqs. (\ref{eq:Friedmann}) and (\ref{eq:FieldMotion}) lead to $3M_{\rm pl}^2 H^2\simeq V$ and $3H\dot{\phi}\simeq -V,_{\phi}$, respectively. The number of e-foldings (Eq. (\ref{eq:efoldings})) can then be written as:
\begin{equation}
N\simeq\frac{1}{M_{\rm pl}^2}\int_{\phi_f}^{\phi}\frac{V}{V,_{\tilde{\phi}}}d\tilde{\phi}, \label{eq:efoldings2}
\end{equation}
where $\phi_f$ is the field value at the end of inflation, known by the relation $\epsilon(\phi_f)=1$.

The slow-roll parameter is expressed as: 
$\epsilon=\dot{\phi}^2/(2M_{\rm pl}^2 H^2)$.
For the slow-roll approximation, it follows that $\epsilon\simeq\epsilon_V$ and $\eta\simeq 4\epsilon_V-2\eta_V$, where
\begin{equation}
\epsilon_V\equiv\frac{M_{\rm pl}^2}{2}\left(\frac{V,_\phi}{V}\right)^2,~~~~~\eta_V\equiv\frac{M_{\rm pl}^2V,_{\phi\phi}}{V}. \label{eq:epsilon_eta}
\end{equation}
Under the fact that $c_s^2=1$, the observables lead to:
\begin{equation}
n_s=1-6\epsilon_V+2\eta_V,~~~~r=-8n_t,~~~~n_t=-2\epsilon_V. \label{eq:ns_r_nt}
\end{equation}
For a given inflaton potential, these observables can be described in terms of $\phi$. The field value that corresponds to $N=50-60$ is determined using Eq. (\ref{eq:efoldings2}).

The tensor-to-scalar ratio $r$ is associated with the variation of the field during inflation. The relation $(d\phi/dN)^2\simeq M_{\rm pl}^2r/8$ is obtained from Eqs. (\ref{eq:efoldings2})-(\ref{eq:ns_r_nt}). Given that $r$ is almost constant during inflation, the field variation $\Delta\phi$ is approximately expressed as:\cite{Lyth1997,Lyth2008}
\begin{equation}
\frac{\Delta\phi}{M_{\rm pl}}\simeq\left(\frac{r}{2\times 10^{-3}}\right)^{1/2}\left(\frac{N}{60}\right). \label{eq:Deltaphi_Mpl}
\end{equation}
The models with $\Delta\phi < M_{\rm pl}$ are called small-field inflation models, when $r$ is smaller than $2\times 10^{-3}$ for $N = 60$. 
Here we adopt the criterion according to Eq. (\ref{eq:Deltaphi_Mpl}) to classify the large-field and small-field models.

Small-field inflation can be achieved by the potential
\begin{equation}
V(\phi)=\Lambda^4[1-\mu(\phi)],
\end{equation}
where $\Lambda$ is a constant and $\mu(\phi)$ is a function of $\phi$. For D-brane inflation \cite{Dvali1999} and K$\ddot{\rm a}$hler-moduli inflation, \cite{Conlon2006} we set $\mu(\phi)=e^{-\phi/M}$ and $\mu(\phi)=c_1\phi^{4/3}e^{-c_2\phi^{4/3}}$ ($c_1>0$, $c_2>0$), respectively; in Refs. \cite{Kachru2003, BlancoPillado2006, Baumann2008, Panda2007} other similar models are studied. For these models, the so-called $\eta$-problem for the natural parameters restricted by string theory must be addressed.
For the function $f(\phi)=e^{-\phi /M}$, the number of e-foldings is described by $N\simeq(M/M_{\rm pl})^2e^{\phi /M}$, in which case $n_s$ and $r$ are
\begin{equation}
n_s\simeq 1-\frac{2}{N}, ~~~~~ r\simeq\frac{8}{N^2}\left(\frac{M}{M_{\rm pl}}\right)^2.
\end{equation}
For $M < M_{\rm pl}$ and $50 < N < 60$, $0.960<n_s<0.967$ and $r <2.6\times10^{-3}$.

For K$\ddot{\rm a}$hler-moduli inflation, $n_s$ and $r$ are in the ranges of $0.960 < n_s < 0.967$ and $r < 10^{-10}$ for $50 < N < 60$ \cite{Conlon2006}, so that the model is classified as very small-field inflation.

$n_s$ and $r$ predicted by the inflation models are summarized in Table \ref{tab:table1}.

\begin{center}
\begin{table}[h!]
\caption{
Predictions for $n_s$ and $r$ from the models. \label{tab:table1}}
\begin{tabular}{l l l}
\hline
Model & Spectral index & Tensor-to-scalar ratio \\
\hline\hline
D-brane & $0.960<n_s<0.967$ & $~~~r<2.6\times10^{-3}$ \\
\hline
K$\ddot{\rm a}$hler-moduli~~ & $0.960<n_s<0.967$ & $~~~r<10^{-10}$ \\
\hline
\end{tabular}
\end{table}
\end{center}

\section{Comparison with Planck 2018 \label{sec:planck}}

In the 2018 release of the Planck CMB anisotropy observations \cite{Akrami2018}, the results of cosmic inflation were consistent with the two previous Planck releases \cite{Ade2014,Ade2016}; however, the uncertainties were smaller due to improvements in the characterization of polarization at low and high multipoles. The spectral index of scalar perturbations obtained from the Planck temperature, polarization, and lensing data was $n_s=0.9649\pm 0.0042$ at 68\% C.L., i.e., there is no evidence for a scale dependence of $n_s$. Spatial flatness is ascertained at a precision of 0.4\% at 95\% C.L. from the combination with baryon acoustic oscillation (BAO) data. The Planck upper limit on the tensor-to-scalar ratio, $r_{0.002}<0.10$ at 95\% C.L., by combination with the BICEP2/Keck Array (BK14) data, is further tightened to obtain $r_{0.002}<0.064$.

\begin{figure}[h!]
\hspace*{10mm}
\includegraphics[width=90mm]{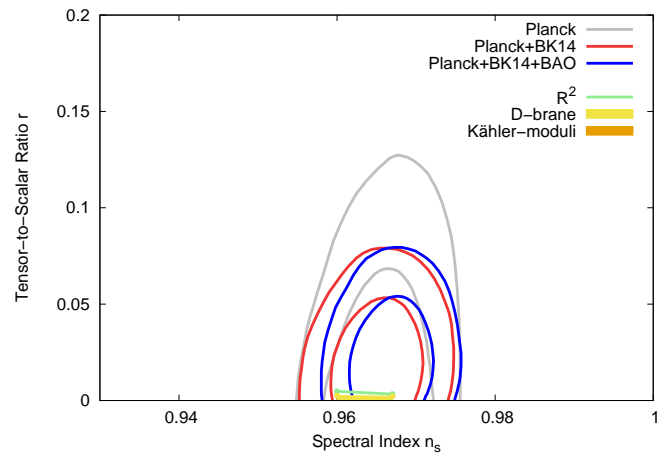}
\vspace{3mm}
\caption{Constraints for $n_s$ and $r$ from Planck 2018 data alone, and in combination with BK14 or BK14 plus BAO data, compared to the theoretical predictions of inflationary models. \label{fig:Dbrane}}
\end{figure}

The ($n_s$, $r$)-planes are plotted in Fig.\ref{fig:Dbrane}. Gray lines are constraints from the Planck 2018 data, and red lines are constraints from the Planck 2018 data combined with the BK14 data, and blue lines are constraints from the Planck 2018 data combined with the BK14 data and that with the addition of the BAO data. The lines show the 1$\sigma$ (68\%) and 2$\sigma$ (95\%) confidence limits, respectively. The yellow region is the prediction by the D-brane inflation model. The orange region is the prediction by the K$\ddot{\rm a}$hler-moduli inflation model. The region predicted by the Starobinsky inflation model is also plotted (green line).

\begin{figure}[h!]
\hspace*{10mm}
\includegraphics[width=90mm]{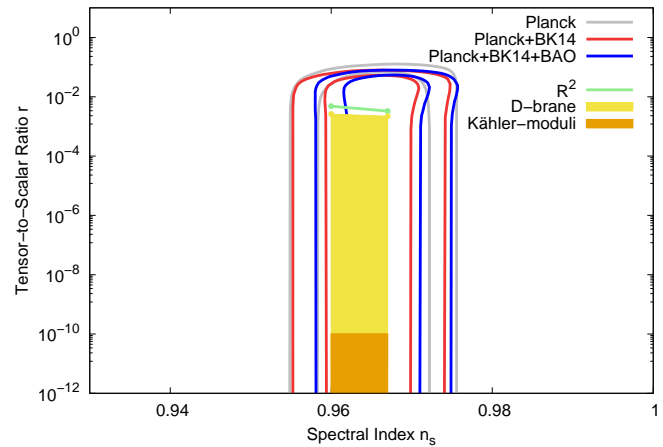}
\vspace{3mm}
\caption{The same as Fig.\ref{fig:Dbrane}, except with the tensor-to-scalar ratio expressed logarithmically. \label{fig:moduli}}
\end{figure}

Fig.\ref{fig:moduli} is the same as Fig.\ref{fig:Dbrane}, except that the tensor-to-scalar ratio is expressed logarithmically.

The D-brane and K$\ddot{\rm a}$hler-moduli inflation models are consistent with the observational data.

\section{Future observations \label{sec:futureOBS}}

Single field models with slow roll conditions give the approximation formula (Lyth relation \cite{Lyth1997,Lyth2008}):
\begin{equation}
r\simeq 0.002\left(\frac{60}{N}\right)^2\left(\frac{\Delta\phi}{M_{\rm pl}}\right)^2.
\end{equation}

Next-generation CMB satellites (LiteBIRD \cite{Hazumi2011}, COrE \cite{Finelli2018}, and PIXIE \cite{Kogut2011}) are being designed to detect the primordial CMB B-mode polarization at large angular scales. One goal of LiteBIRD is to reach the total uncertainty of $r$, $\delta r=0.001$. When this is achieved, the non-zero tensor-to-scalar ratio $r$ is detected, which has more than 10$\sigma$ significance for $r > 0.01$. If the primordial B-mode is not detected by next-generation satellites, an upper limit of $r < 0.002$ (95\% C.L.) will be obtained, which means that all the large-field models classified by the Lyth relation will be rejected. Therefore, inflation models that predict very small values of $r$ must be investigated. Furthermore, the effect of the quantum fluctuation on the value of $r$ may be partial; therefore, it is important to consider part of the quantum fluctuation origin.

\section{Summary \label{sec:summary}}

Inflation models that predict very small values of $r$ were investigated. We compare the constraints by the Planck 2018 data combined with other CMB data, and cosmological observations with the predictions of the inflation models with respect to $n_s$ and $r$. The D-brane and K$\ddot{\rm a}$hler-moduli inflation models were consistent with the observational data.

Furthermore, we discuss the comparison of future tensor-to-scalar ratio data with the predictions from the inflation models. One of the goals of a next-generation CMB satellite (LiteBIRD) is to reach the total uncertainty of the tensor-to-scalar ratio $r$, $\delta r=0.001$. If the primordial B-mode is not detected by next-generation satellites, an upper limit $r < 0.002$ (95\% C.L.) will be obtained, which means that all the large-field models classified by the Lyth relation will be rejected. In this case, inflation models such as the K$\ddot{\rm a}$hler-moduli model will survive. Therefore, the study of inflation models that predict the very small value of the tensor-to-scalar ratio $r$, is necessary. Furthermore, the effect of quantum fluctuation on the value of $r$ may be partial; therefore, it is important to consider part of the quantum fluctuation origin.

\bibliography{hirano.bib}
\end{document}